\documentclass[a4paper,11pt]{article}
\pdfoutput=1
\usepackage{jcappub}
\usepackage{pifont}

\graphicspath{{plots/}}

\newcommand{\gs}{g_\text{eff}}
\newcommand{\gss}{h_\text{eff}}
\newcommand{\Trh}{T_\text{rh}}
\newcommand{\Tmax}{T_\text{max}}
\newcommand{\sv}{\langle\sigma v\rangle}
\newcommand{\rR}{\rho_R}
\newcommand{\rp}{\rho_\phi}
\newcommand{\Gp}{\Gamma_\phi}
\newcommand{\mzp}{m_{Z'}}
\newcommand{\vf}{V_f}
\newcommand{\mdm}{m_\chi}
\newcommand{\vdm}{V_\chi}

\title{$Z'$-Mediated Dark Matter with Low-Temperature Reheating}
\author[a]{Geneviève Bélanger,}
\author[b]{Nicolás Bernal,}
\author[c]{Alexander Pukhov}

\affiliation[a]{LAPTh, CNRS, Université Savoie Mont-Blanc\\
9 Chemin de Bellevue, 74940 Annecy, France}
\affiliation[b]{New York University Abu Dhabi\\
PO Box 129188, Saadiyat Island, Abu Dhabi, United Arab Emirates}
\affiliation[c]{Skobeltsyn Institute of Nuclear Physics, Moscow State University\\
Moscow 119992, Russia}

\emailAdd{belanger@lapth.cnrs.fr}
\emailAdd{nicolas.bernal@nyu.edu}
\emailAdd{alexander.pukhov@gmail.com}

\abstract{We consider a simple extension of the standard model with fermionic dark matter (DM) and a $Z'$ gauge boson acting as a mediator. We also assume a scenario where cosmic reheating occurs at low temperatures due to the decay of a massive inflaton into standard model states. To follow the evolution of the background and the dark sector states, we implement the required Boltzmann equations in the code micrOMEGAs to explore both the freeze-out and freeze-in  mechanisms. We determine the parameter space of the model that satisfies the relic density constraint under different assumptions for the reheating dynamics, and examine current constraints from DM direct detection, taking special care of the scenarios where DM was produced during the reheating era. Large regions of the parameter space favored by low-temperature reheating cases are already probed or will be within the reach of future experiments, both for the WIMP and the FIMP paradigms.}

\begin{document}
\begin{flushright}
\end{flushright}
\maketitle

\section{Introduction}
Although the evidence for dark matter (DM) is very convincing, its nature remains a mystery. Many possibilities have been entertained for a new DM particle covering a wide range of masses and couplings~\cite{Bertone:2004pz, Cirelli:2024ssz}. To ascertain the viability of a DM model, one of the most important criteria is that the prediction of the DM relic density matches the value determined by PLANCK~\cite{Planck:2018vyg}. The calculation of the relic density is usually performed in the standard cosmological scenario, where it is assumed that standard model (SM) radiation dominated the energy density of the universe right from the end of inflationary reheating until the onset of Big Bang nucleosynthesis (BBN). Moreover, it is assumed that the reheating temperature, which is defined as the beginning of the radiation-dominated (RD) era, is larger than the temperature at which the DM is produced. Many studies have shown that departure from the standard cosmological scenario would significantly affect the prediction for the relic density of DM, for example through a modification of the entropy in the early universe and/or by considering a low reheating temperature $\Trh$~\cite{Giudice:2000ex, Fornengo:2002db, Pallis:2004yy, Gelmini:2006pw, Drees:2006vh, Yaguna:2011ei, Roszkowski:2014lga, Drees:2017iod, Bernal:2018ins, Bernal:2018kcw, Cosme:2020mck, Arias:2021rer, Bernal:2022wck, Bhattiprolu:2022sdd, Haque:2023yra, Ghosh:2023tyz, Silva-Malpartida:2023yks, Arias:2023wyg, Bernal:2024yhu, Silva-Malpartida:2024emu, Bernal:2024ndy}. In fact, the reheating temperature is not strongly constrained and can generally take any value above a few MeVs~\cite{Sarkar:1995dd, Kawasaki:2000en, Hannestad:2004px, DeBernardis:2008zz, deSalas:2015glj}. The presence of a scalar field, the inflaton, that decays into SM radiation during reheating leads to an increase in the entropy of the Universe during the reheating phase~\cite{Allahverdi:2020bys, Batell:2024dsi}. The decay width of the inflaton will also determine the value of the reheating temperature, a narrow width being associated with a low value for $\Trh$. 

Injection of entropy during the reheating phase can reduce by several orders of magnitude the required thermally averaged cross-section for WIMPs~\cite{Gelmini:2006pw, Bernal:2022wck, Bernal:2024yhu}. WIMPs having reduced coupling to the SM, would hence more easily escape detection in current collider and astroparticle searches. In models where DM is a feebly interacting particle, the entropy injection implies an increase in the DM production cross-section through the freeze-in mechanism, thus increasing the coupling required to satisfy the relic density constraint~\cite{Silva-Malpartida:2023yks, Silva-Malpartida:2024emu}. In addition, many studies have shown that reducing the reheating temperature below the typical scale for freeze-in ($T\approx M$) where $M$ is the mass of DM or of the mediator that decays into FIMPs also increases the typical couplings required to reproduce the relic density~\cite{Belanger:2018sti, Ahmed:2022tfm, Cosme:2023xpa, Gan:2023jbs, Cosme:2024ndc, Arcadi:2024wwg, Boddy:2024vgt, Arcadi:2024obp, Lebedev:2024vor, Bernal:2024ndy}.

In this paper, we consider a scenario in which an inflaton slowly decays into SM radiation, and we take the width of this inflaton as a free parameter.  We implement the resulting modifications to the Boltzmann equations in the code micrOMEGAs~\cite{Alguero:2023zol} to allow to compute the DM relic density for both weakly or feebly interacting particles. To illustrate the impact of the inflaton decays on DM observables, we consider a simple extension of the SM with a $Z'$ gauge boson and a fermion DM. We determine the parameter space of the model that satisfies the relic density constraint under different assumptions for the inflaton decay width and examine current and future constraints from DM direct detection (DD). We also determine the condition for DM to behave as a WIMP or as a FIMP.

The paper is organized as follows. Section~\ref{sec:LTR} describes the Boltzmann equations, including inflaton decays. Section~\ref{sec:DM} derives numerical results in the framework of a simplified model with a $Z'$ and a fermionic DM. Section~\ref{sec:conclu} contains our conclusion. The appendix~\ref{sec:appendix} contains a description of the new functions of micrOMEGAs that allow one to compute the relic density in the presence of an inflaton. 

\section{Low-Temperature Reheating} \label{sec:LTR}
During cosmic reheating, the inflaton, $\phi$, is assumed to decay into SM radiation with a total decay width $\Gp$. The dynamics of the background is driven by the set of Boltzmann equations for the inflaton energy density, $\rp$, and the SM entropy density, $s$~\cite{Gelmini:2006pw}
\begin{align}
    \frac{d\rp}{dt} + 3\, H\, \rp &= - \Gp\, \rp\,, \label{eq:BEa0}\\
    \frac{ds}{dt} + 3\, H\, s &= + \frac{\Gp\, \rp}{T}\,, \label{eq:BEb0}
\end{align}
where the entropy density is defined as a function of the SM temperature, $T$, as
\begin{equation}
    s(T) = \frac{2\pi^2}{45}\, \gss(T)\, T^3,
\end{equation}
with $\gss(T)$ being the number of relativistic degrees of freedom contributing to $s$~\cite{Drees:2015exa}. Besides the source term, Eqs.~\eqref{eq:BEa0} and~\eqref{eq:BEb0} are coupled by the Friedmann equation for the Hubble expansion rate $H$ given by
\begin{equation}
    H^2 = \frac{8\pi}{3}\, \frac{\rp + \rR}{M_P^2} \,,
\end{equation}
where $M_P \simeq 1.2 \times 10^{19}$~GeV is the Planck mass, and the SM energy density, $\rR$, is given by
\begin{equation}
    \rR(T) = \frac{\pi^2}{30}\, \gs(T)\, T^4,
\end{equation}
with $\gs(T)$ is the number of relativistic degrees of freedom contributing to $\rR$.

\begin{figure}[t!]
    \def\sepf{0.496}
    \centering
    \includegraphics[width=\sepf\columnwidth]{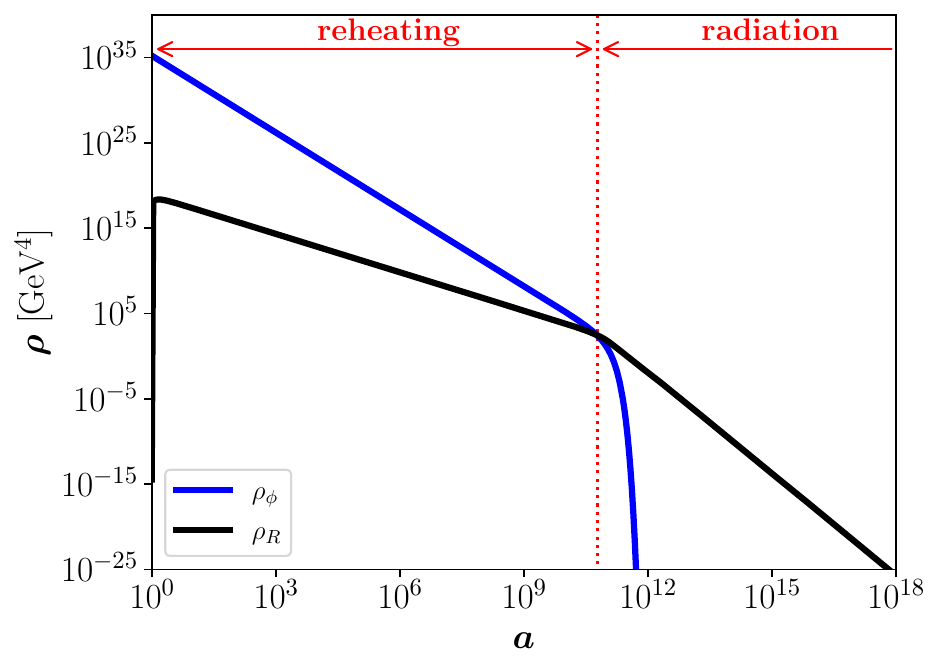}
    \includegraphics[width=\sepf\columnwidth]{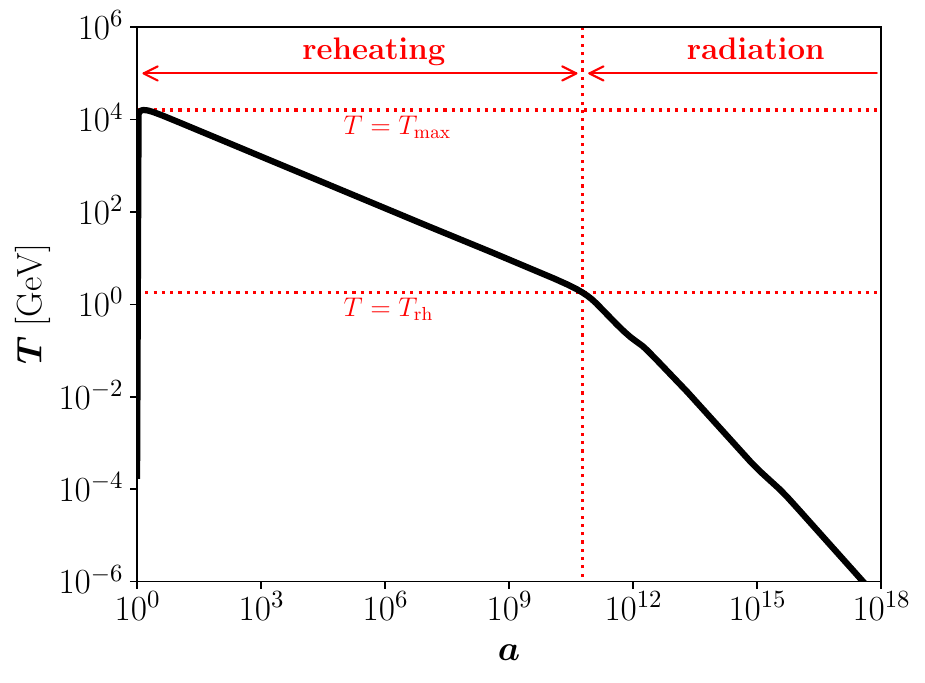}
    \caption{Left: Evolution of the energy densities for the inflaton and the SM radiation as a function of the scale factor $a$. Right: Evolution of the SM temperature. In both panels $H_I = 10^{-1}$~GeV and $\Gp = 10^{-17}$~GeV were used. The vertical lines show the boundary between the reheating and the SM radiation dominated era, while the horizontal lines correspond to $T = \Tmax$ and  $T = \Trh$.}
    \label{fig:evolution}
\end{figure}

To solve Eqs.~\eqref{eq:BEa0} and~\eqref{eq:BEb0}, it is convenient to rewrite them as a function of $z_\phi \equiv \rho_\phi \times a^3$, $z_s \equiv s^{4/3} \times a^4$, and the cosmic scale factor $a$ as
\begin{align} 
    \label{Infl_evol_rho} 
    \frac{dz_\phi}{da} &= -\frac{\Gp}{H\, a}\, z_\phi\,,\\
    \label{Infl_evol_s} 
    \frac{dz_s}{da} &= +\frac43\, \frac{\Gp}{H}\, \frac{s^{1/3}}{T}\, z_\phi\,,
\end{align}
with the initial conditions at the beginning of reheating $t = t_I$ (corresponding to the scale factor $a_I \equiv a(t_I) = 1$) $\rR(t_I) = 0$ and $\rp(t_I) = \frac{3}{8\pi}\, M_P^2\, H_I^2 \ne 0$, where $H_I$ corresponds to the inflationary scale. Interestingly, non-observation of B modes implies an upper bound on the inflationary scale $H_I < 4.0 \times 10^{-6}~M_P$~\cite{BICEP:2021xfz}. The left panel of Fig.~\ref{fig:evolution} shows a fully numerical solution for the evolution of the energy densities as a function of the scale factor, for $H_I = 10^{-1}$~GeV and $\Gp = 10^{-17}$~GeV, giving rise to $\Trh \simeq 1.8$~GeV. During reheating $\rp(a) \propto a^{-3}$ while $\rR(a) \propto a^{-3/2}$. After the onset of the radiation-dominated era, $\phi$ decays exponentially fast, and $\rR(a) \propto a^{-4}$. The corresponding evolution of the SM temperature is shown in the right panel of Fig.~\ref{fig:evolution}. During reheating, as the SM radiation is not free but sourced by the inflaton, $T(a) \propto a^{-3/8}$; after reheating, the standard scaling $T(a) \propto a^{-1}$ is recovered.

The end of reheating is defined as the onset of SM radiation domination, at a temperature $T = \Trh$ defined as $\rR(\Trh) = \rp(\Trh)$. To avoid spoiling the success of BBN, $\Trh > T_\text{BBN} \simeq 4$~MeV~\cite{Sarkar:1995dd, Kawasaki:2000en, Hannestad:2004px, DeBernardis:2008zz, deSalas:2015glj} is required. The reheating temperature can also be estimated implicitly by the equality $H(\Trh) = \Gp$, which implies
\begin{equation}
    \Trh^2 = \frac{3}{2\pi} \sqrt{\frac{5}{\pi\, \gs(\Trh)}}\, M_P\, \Gp\,.
\end{equation}
Additionally, the maximum temperature $\Tmax$ reached during reheating by the thermal bath can be estimated by~\cite{Barman:2021ugy}
\begin{equation}
    \Tmax^4 = \frac{15}{2 \pi^3\, \gs(\Tmax)} \left(\frac38\right)^\frac85 M_P^2\, \Gp\, H_I\,.
\end{equation}
In the present analysis, the solution of the Boltzmann equations and the determination of $\Trh$ and $\Tmax$ is performed numerically using the code micrOMEGAs~\cite{Alguero:2023zol}, for a description of the new routines, see Appendix~\ref{sec:appendix}.

\section{\boldmath $Z'$-Mediated Dark Matter} \label{sec:DM}
We assume that DM is a Dirac fermion,  $\chi$, with mass $\mdm$ that couples to the SM through a $Z'$ mediator of mass $\mzp$. We further assume {\it vectorial} couplings of the $Z'$ to the DM and SM particles to avoid anomalies in the axial vector case. The couplings are $\vdm$ and $\vf$, respectively. The extension of the SM Lagrangian includes the terms
\begin{equation}
    \mathcal{L} \supset - \mdm\, \bar\chi\chi - \frac12\, \mzp\, Z'_\mu Z^{\prime\mu} + \vdm\, \bar\chi\, \gamma^\mu\,\chi\, Z'_\mu + \sum_f \vf\, \bar f\, \gamma^\mu\, f\, Z'_\mu\,,
\end{equation}
where $f$ correspond to the SM fermionic fields.

The evolution of the DM number density $n$ can be tracked by the Boltzmann equation
\begin{equation} \label{eq:BEc0}
    \frac{dn}{dt} + 3\, H\, n = - \langle\sigma v\rangle \left(n^2 - n_\text{eq}^2\right),
\end{equation}
where $\langle\sigma v\rangle$ is the thermally averaged DM annihilation cross section and $n_\text{eq}$ is the DM number density at equilibrium, which in the non-relativistic limit reduces to
\begin{equation}
    n_\text{eq}(T) = 2 \left(\frac{\mdm\, T}{2\pi}\right)^\frac32 e^{-\frac{\mdm}{T}}.
\end{equation}
Equation~\eqref{eq:BEc0} can be conveniently rewritten as a function of $z_\chi= n\, a^3$ as 
\begin{equation} \label{eq:BEc1} 
    \frac{dz_\chi}{da} = -\frac{1}{H a^4}\, \sv \left(z_\chi^2 -\bar{z}_\chi^2\right),
\end{equation}
where $\bar{z}_\chi \equiv n_\text{eq} \times a^3$.

From a technical perspective, we note that before solving Eq.~\eqref{eq:BEc1} the code tries to find a low-temperature region, where the DM is in thermal equilibrium with the SM bath. We use the following conditions of equilibrium 
\begin{align}
    |\delta z_\chi| &< 10^{-2}\, \bar{z}_\chi\,,\\
    \frac{\delta a}{a} &< 10^{-3},
\end{align}
where $\delta z$ and $\delta a$ are estimations of the deviation from equilibrium and  for the step of integration
\begin{align}
    \delta z_\chi &= -\frac{ d \log(\bar{z}_\chi)}{da}\, \frac{H\, a^4}{2\, \sv}\,,\\
    \delta a &=  \frac{H\, a^4}{\bar{z}_\chi\,  2\, \sv}\,.
\end{align}   
If the conditions for thermal equilibrium are satisfied, the integration starts from the largest scale parameter $a$ where the equilibrium condition holds and the initial value is taken to be $z_\chi=\bar{z}_\chi + \delta z_\chi$. Otherwise, the integration starts from $a=1$ with zero initial condition. With this procedure, it is determined whether the DM reaches or does not reach chemical equilibrium with the SM bath, and therefore, whether the DM is a WIMP or a FIMP relic.

\begin{figure}[t!]
    \def\sepf{0.496}
    \centering
    \includegraphics[width=\sepf\columnwidth]{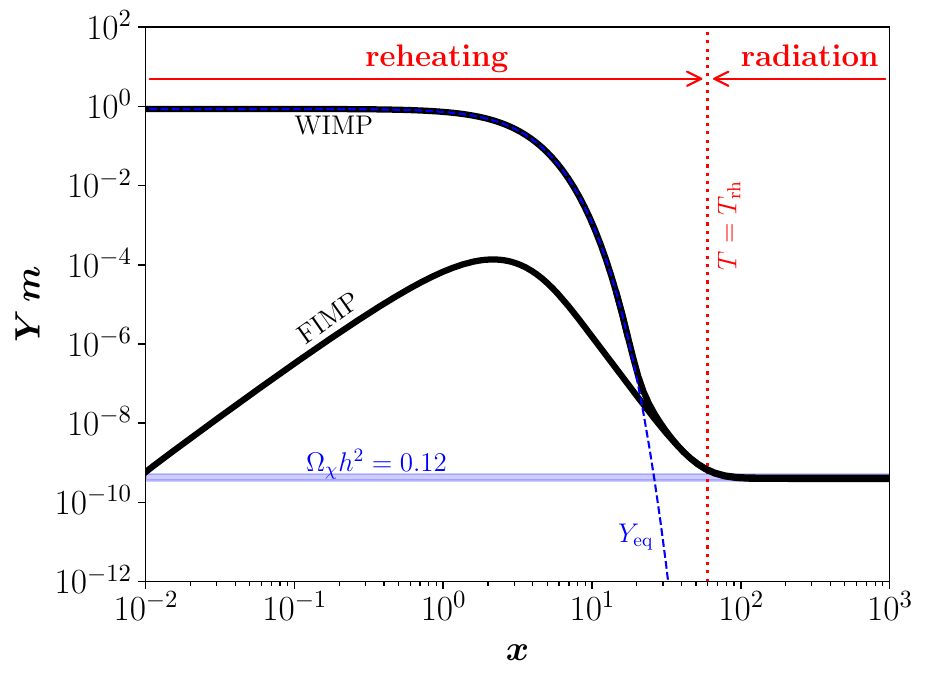}
    \includegraphics[width=\sepf\columnwidth]{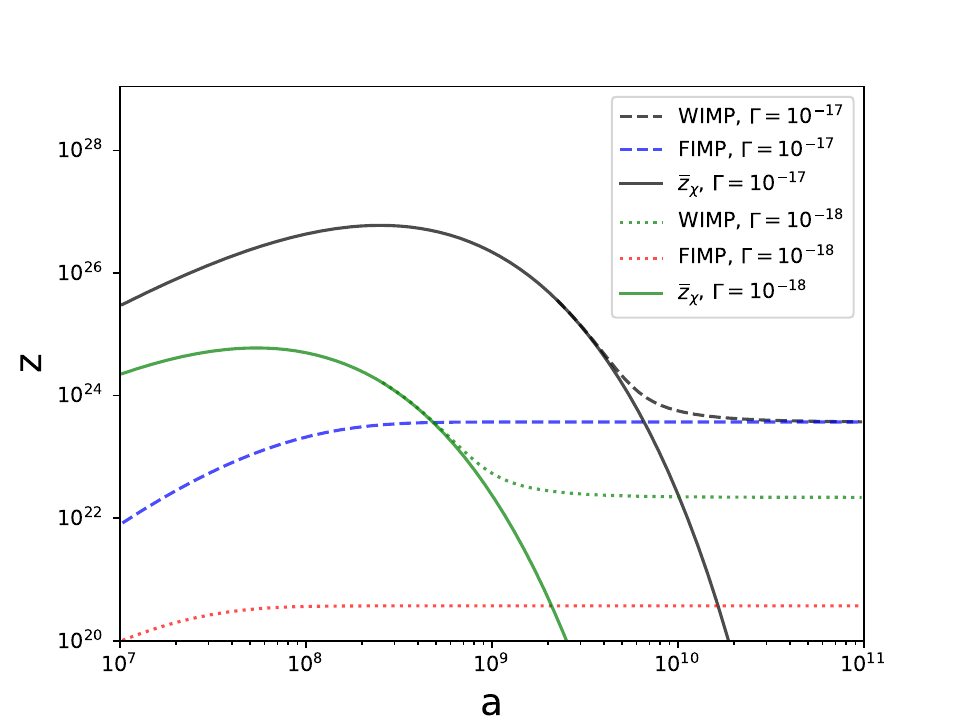}
    \caption{ Evolution of the DM yield $Y\, m = m_\chi\, n_\chi/s$ as a function of $x \equiv \mdm/T$, (left) and evolution of $z_\chi$ as a function of $a$ (right) for $\mdm = 100$~GeV, $H_I = 10^{-1}$~GeV and $\Gp = 10^{-17}$~GeV, $\mzp = 1$~GeV and $\vf = 10^{-5}$. The entire observed DM relic abundance is fitted with $\vdm = 1.1 \times 10^{-1}$ (WIMP solution) or $\vdm \simeq 5.7 \times 10^{-4}$ (FIMP solution). The right plot also shows the evolution of $z_\chi$  for $\Gp = 10^{-18}$~GeV keeping all other parameters the same.}
    \label{fig:yield}
\end{figure}

Equation~\eqref{eq:BEc1} has to be solved in a background defined by Eqs.~\eqref{Infl_evol_rho} and~\eqref{Infl_evol_s}, assuming a vanishing initial DM number density after inflation. Figure~\ref{fig:yield} shows an example of the evolution of DM yield $Y \equiv n/s$ as a function of $x \equiv \mdm/T$, for $\mdm = 100$~GeV, $H_I = 10^{-1}$~GeV, $\Gp = 10^{-17}$~GeV, $\mzp = 1$~GeV and $\vf = 10^{-5}$. The entire observed DM relic abundance is fitted with $\vdm = 1.1 \times 10^{-1}$ (WIMP solution) or $\vdm \simeq 5.7 \times 10^{-4}$ (FIMP solution). Contrary to the behavior in the RD era, after chemical freeze-out, the comoving number density for WIMPs is diluted due to injection of entropy, until $T \sim  \Trh$. The same dilution occurs for FIMPs after their freeze-in. Both features can be observed in Fig.~\ref{fig:yield}. Figure~\ref{fig:evolution} (right) displays the evolution of $z_\chi$  with the scale factor $a$ for both the WIMP and FIMP solutions  for the set of parameters that fit the observed DM relic abundance for $\Gp=10^{-17}$~GeV as mentioned above.  The figure also shows how $z_\chi$ and $\overline{z}_\chi$ decrease for a smaller inflaton width, $\Gp=10^{-18}$~GeV, here masses and couplings are fixed to the previous values and   both the WIMP and the FIMP are underabundant. Note that  $z_\chi$ cannot exceed the maximal value of $\overline{z}_\chi$, which depends on  $\Gp$ and $\mdm$, hence for a fixed  $\Gp$ there is a maximal value  of $\mdm$ that can reproduce the observed relic abundance for the WIMP,  similarly  for a fixed value of $\mdm$  there is a minimal  value  of $\Gp$ which is compatible with $\Omega h^2=0.12$.

For the following phenomenological analysis, we choose two representative benchmark points for $Z'$ mass and coupling to DM:
\begin{itemize}
    \item[\ding{192}] $\mzp = 1$~GeV and $\vf = 10^{-5}$,
    \item[\ding{193}] $\mzp = 500$~GeV and $\vf = 10^{-4}$.
\end{itemize}
For such values of the coupling, $Z'$ escapes current collider searches, see Ref.~\cite{Cosme:2021baj}. Furthermore, we have numerically checked that for these choices, $Z'$ is always in equilibrium with the SM thermal bath; this allows us to avoid solving an extra Boltzmann equation for the $Z'$ density.

The evolution of the cosmological background is defined by $H_I$ and $\Gp$, or alternatively by $\Tmax$ and $\Trh$. Here we assume a large inflationary scale, so that $\Tmax$ is much higher than the scales at which DM is produced and therefore decouples from the low-energy setup. In practice, we set it to $H_I = 0.1$~GeV.

\begin{figure}[t!]
    \def\sepf{0.496}
    \centering
    \includegraphics[width=\sepf\columnwidth]{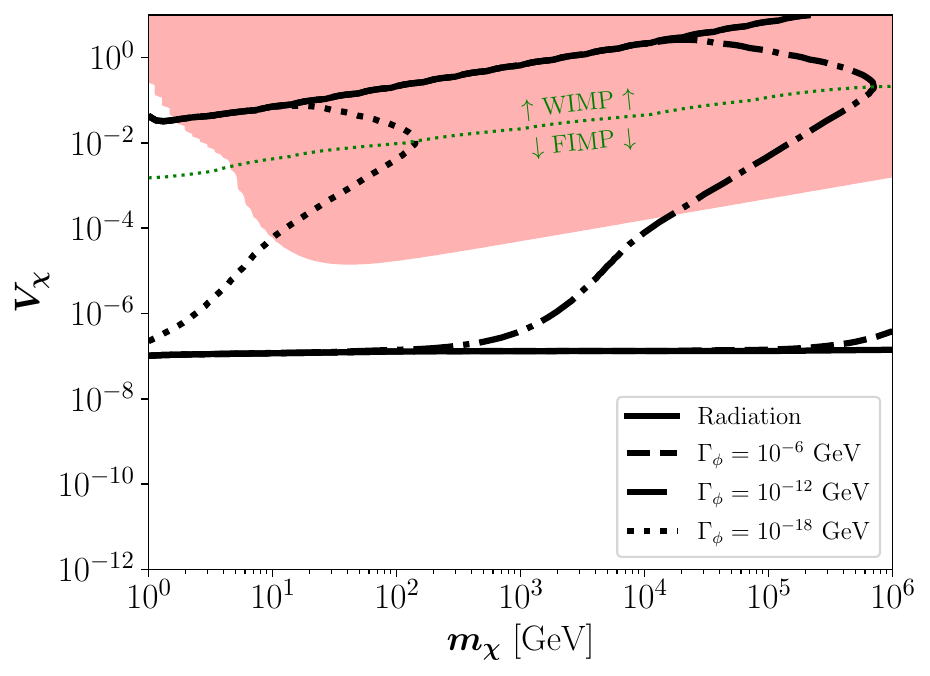}
    \includegraphics[width=\sepf\columnwidth]{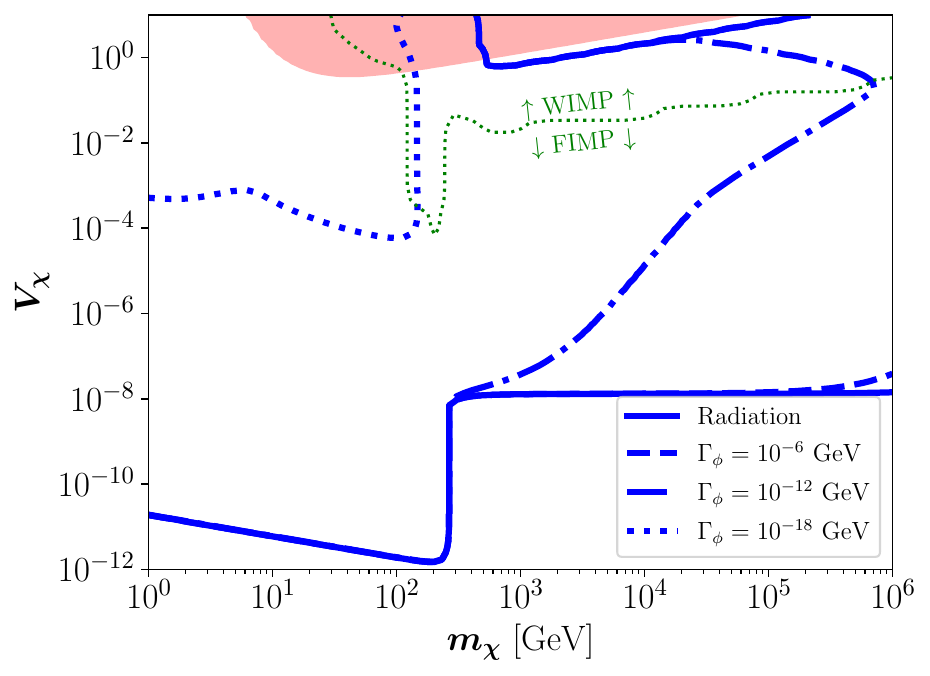}
    \caption{Parameter space that fits the entire observed DM abundance, for the benchmarks \ding{192} ($\mzp = 1$~GeV and $\vf = 10^{-5}$ - left), and \ding{193} ($\mzp = 500$~GeV and $\vf = 10^{-4}$ - right).}
    \label{fig:m-g}
\end{figure} 
Figure~\ref{fig:m-g} shows the parameter space that fits the entire observed DM abundance, for the two benchmark points discussed (\ding{192} left and \ding{193} right). We consider three values of the inflaton width, $\Gp = 10^{-6}$~GeV, $\Gp = 10^{-12}$~GeV and $\Gp = 10^{-18}$~GeV, which correspond to $\Trh \simeq 9.7 \times 10^4$~GeV, $\Trh \simeq 99$~GeV and $\Trh \simeq 0.15$~GeV, respectively (and to $\Tmax \simeq 9.1 \times 10^6$~GeV, $\Tmax \simeq 2.9 \times 10^5$~GeV, and $\Tmax \simeq 9.1 \times 10^3$~GeV, respectively). The left panel of Fig.~\ref{fig:m-g} represents the point \ding{192} with a GeV-scale mediator. For high-temperature reheating (that is, high values of $\Gp \gg 10^{-6}$~GeV), the two solid lines with $\vdm  \sim \mathcal{O}(10^0)$ and $\vdm  \sim \mathcal{O}(10^{-7})$ correspond to the usual WIMP and FIMP solutions in radiation domination, respectively. In this case where $Z'$ is lighter than DM and its coupling to SM is suppressed ($\vf = 10^{-5}$), the process responsible for DM production is pair annihilation of $Z'$ into pairs of DM or the reverse process. The parameter space between the WIMP and FIMP solutions overproduces DM and is therefore ruled out. However, a reduction in $\Trh$ (driven by a small $\Gp$) implies that the genesis of DM can occur during the reheating era and some interesting phenomenological consequences appear: $i)$ The parameter space previously excluded between the WIMP and FIMP solutions in RD becomes viable, as the WIMP mechanism can be realized for smaller values of $\vdm$, while FIMPs can occur with larger values of $\vdm$. These effects come from the large entropy injection from inflaton decays. $ii)$~The dotted green line shows the minimal value of the coupling required to achieve chemical equilibrium. Above that line, DM reaches chemical equilibrium with the SM bath and therefore corresponds to a WIMP solution. Alternatively, below the line, DM never reaches chemical equilibrium with the SM bath and is therefore a FIMP. $iii)$ Low-temperature reheating allows a smooth transition between otherwise disconnected WIMP and FIMP solutions. $iv)$ Higher WIMP masses, beyond the usual limit $\mdm \lesssim 130$~TeV from unitarity~\cite{Griest:1989wd}, become accessible~\cite{Bhatia:2020itt, Bernal:2023ura}. $v)$ The current limit of DD experiments, namely LZ~\cite{LZCollaboration:2024lux}, excludes most of the parameter space of the WIMP solution and part of the parameter space for the FIMP solution when the inflaton width is small. FIMPs will be further probed with next-generation experiments, such as DARWIN~\cite{DARWIN:2016hyl}. Note that the limits are particularly strong since the mediator is light.

The right panel of Fig.~\ref{fig:m-g} corresponds to the benchmark \ding{193}, a scenario with a heavier mediator. The solid blue line at $\vdm  \sim \mathcal{O}(10^0)$ shows the WIMP solution in radiation domination, where DM annihilates in a pair of $Z'$. This channel kinematically closes when $\mdm \sim \mzp$, implying an increase in the coupling $\vdm$. WIMP DM cannot be realized for $\mdm < \mzp$, as the coupling $\vf$ is very suppressed ($\vf = 10^{-4}$). However, the FIMP solution in radiation domination (that is, the lower thick blue line) for $\vdm  \sim \mathcal{O}(10^{-9})$ is viable for a wide range of DM masses. In particular, if $\mdm \lesssim \mzp/2$, DM is mainly produced in reactions mediated by an on-shell $Z'$ (or equivalently, by decays of $Z'$ in thermal equilibrium with the SM bath). In contrast, if $\mdm \gtrsim \mzp/2$, DM is mainly generated from pair annihilations of $Z'$ bosons. Here again, low-temperature reheating decreases (increases) the couplings required for the WIMP (FIMP) mechanism. Interestingly, for the case $\Gp = 10^{-18}$~GeV, light WIMPs with masses $\mdm < \mzp/2$ become accessible for $\vdm \gtrsim 10^{-3}$~GeV, this is because DM annihilation is enhanced by the $Z'$ resonance. When the DM mass drops below $\sim 140$~GeV, the resonance effect diminishes and a WIMP solution cannot be reached even for $\vdm \gtrsim 1$. Moreover, for this value of $\Gp$, a rather large coupling to DM, $\vdm \simeq \mathcal{O}(10^{-4})$, is found for the FIMP solution. Note that the current DD limits constrain only the WIMP region when the coupling $\vdm \simeq {\cal O}(1)$, while the next generation of experiments such as DARWIN will increase the sensitivity on the coupling by about one order of magnitude, not enough to probe the FIMP region.

\begin{figure}[t!]
    \def\sepf{0.496}
    \centering
    \includegraphics[width=\sepf\columnwidth]{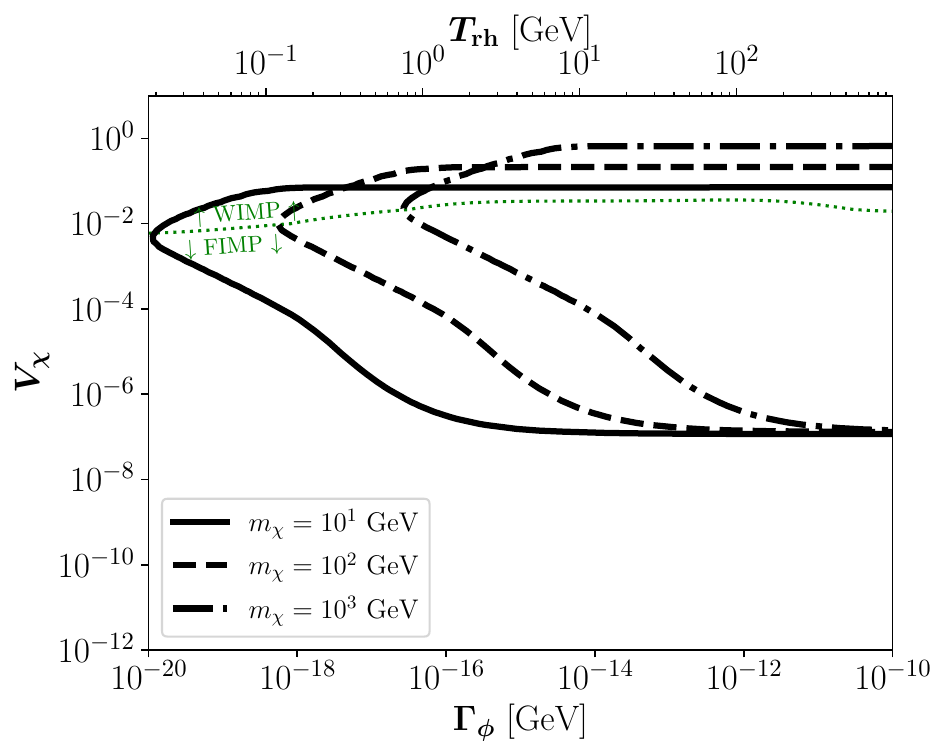}
    \includegraphics[width=\sepf\columnwidth]{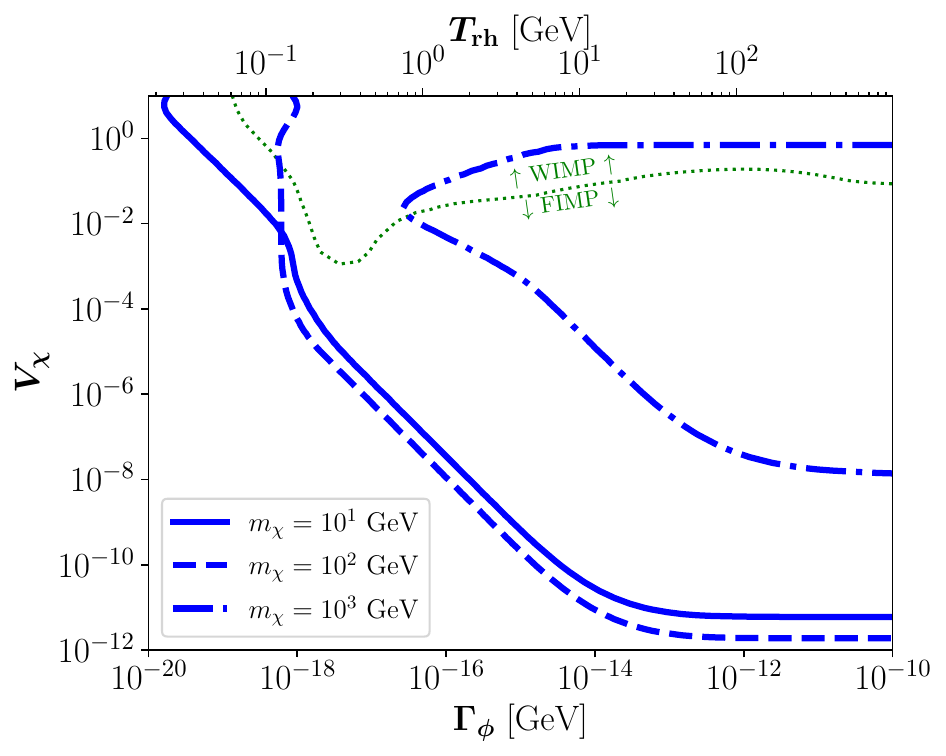}
    \caption{Parameter space that fits the entire observed DM abundance, for the benchmarks \ding{192} ($\mzp = 1$~GeV and $\vf = 10^{-5}$ - left), and \ding{193} ($\mzp = 500$~GeV and $\vf = 10^{-4}$ - right).}
    \label{fig:G-g}
\end{figure} 
The couplings required to fit the observed relic abundance for different reheating temperatures are shown in Fig.~\ref{fig:G-g}, for the benchmarks \ding{192} (left) and \ding{193} (right), and different values of the DM mass. For high-temperature reheating (that is, $\Trh \gg \mdm$), DM is produced in the standard RD era and, therefore, its abundance is independent of $\Trh$. In the left panel, the WIMP and FIMP solutions in RD correspond to horizontal lines with $\vdm \sim \mathcal{O}(10^0)$ and $\vdm \lesssim \mathcal{O}(10^{-7})$, respectively. Additionally, here it can be clearly seen that the WIMP and FIMP solutions merge during reheating because of the decrease (increase) of the required couplings for WIMPs (FIMPs). For a given DM mass, a minimal value can be set for $\Gp$ (or equivalently, for $\Trh$) to match the observed abundance of DM. However, the WIMP region is excluded from the DD limits, as shown in Fig.~\ref{fig:m-g}. The right panel of Fig.~\ref{fig:G-g} shows a similar behavior, with the notable difference that the WIMP solution cannot be easily realized if $\mdm < \mzp$, as expected from Fig.~\ref{fig:m-g}. It is also interesting to note that even for $\vdm \sim \mathcal{O}(1)$ the FIMP solution is viable.

\begin{figure}[t!]
    \def\sepf{0.496}
    \centering
    \includegraphics[width=\sepf\columnwidth]{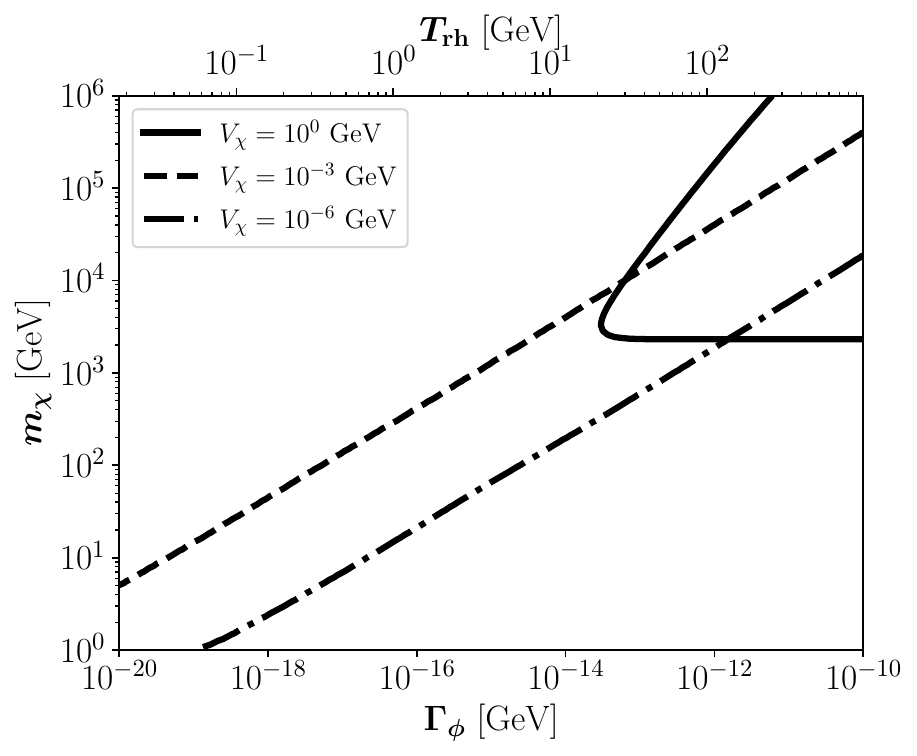}
    \includegraphics[width=\sepf\columnwidth]{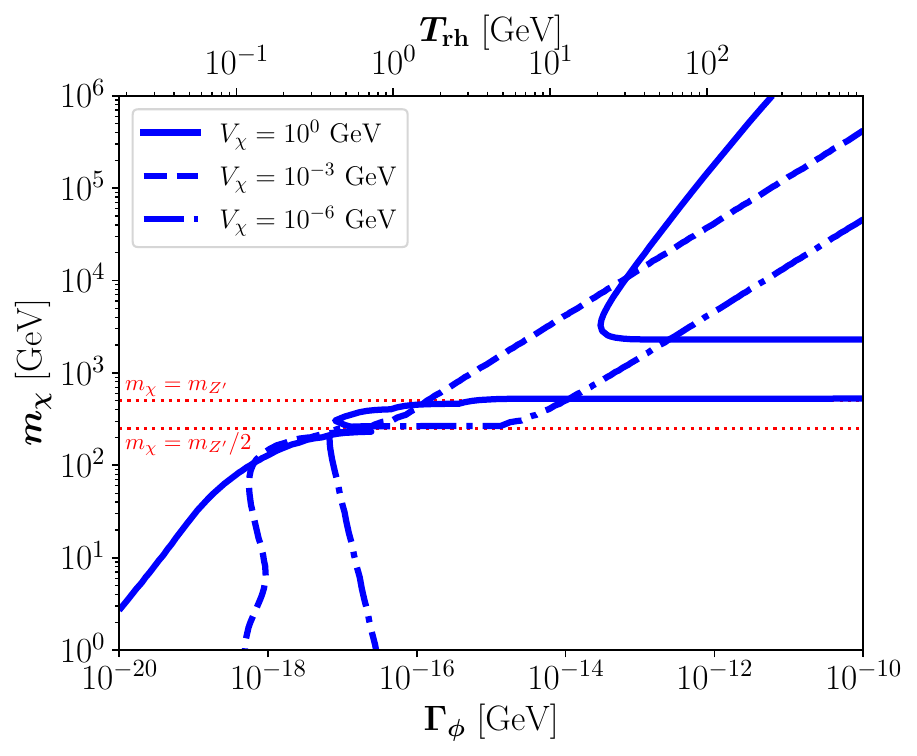}
    \caption{Parameter space that fits the entire observed DM abundance, for the benchmarks \ding{192} ($\mzp = 1$~GeV and $\vf = 10^{-5}$ - left), and \ding{193} ($\mzp = 500$~GeV and $\vf = 10^{-4}$ - right).}
    \label{fig:G-m}
\end{figure} 
Figure~\ref{fig:G-m} shows the parameter space that fits the entire observed DM abundance in the plane $[\Gp,\, \mdm]$, for the two benchmarks discussed. In the left panel, corresponding to the point \ding{192}, only the case $\vdm = 10^0$ (solid line) generates WIMPs: in the horizontal branch at $\mdm \simeq 2$~TeV DM is produced in the RD era, while in the tilted branch during reheating. Note that DM is overabundant in the area between the two branches. The smaller couplings ($\vdm = 10^{-3}$ and $\vdm = 10^{-6}$) are too small to bring the DM to chemical equilibrium and therefore correspond to FIMP and in this case to production during reheating. For both couplings, all masses below the given contours lead to a DM overproduction.

The right panel of Fig.~\ref{fig:G-m} (corresponding to the benchmark \ding{193}) features similarities to the left panel, although there are important differences. For $\vdm = 10^0$, a new WIMP solution appears for $\mdm < \mzp$ where DM is produced by annihilations of $Z'$ pairs. Decays of $Z'$ also contribute if $\mdm < \mzp/2$. Near $\mdm = \mzp/2$, DM annihilates resonantly through the $s$-channel exchange of an on-shell $Z'$ boson, which has to be compensated for by an increase of the required value for $\Trh$. For lower couplings ($\vdm = 10^{-6}$), DM mostly behaves like a FIMP. Note that the production rate increases when $\mdm < \mzp/2$ due to the kinematical opening of the channel $Z' \to \chi \bar\chi$, and hence $\Trh$ has to increase. For the intermediate coupling ($\vdm = 10^{-3}$), a similar behavior is observed with the exception of a small region when the DM mass is in the range $[140,\, 250]$~GeV, where the DM annihilation is enhanced by the $Z'$-resonance and is large enough for the DM to freeze-out. For lighter masses, DM behaves like a FIMP. These features are consistent with the WIMP/FIMP regions shown in Fig.~\ref{fig:m-g}.

\section{Conclusions} \label{sec:conclu}
The understanding of the dark matter (DM) genesis in the early Universe requires having under control not only the particle-physics dynamics but also the evolution of the cosmological background. The history of the primordial Universe before the Big Bang nucleosynthesis era is uncertain. However, it is expected that after inflation, a reheating era takes place, where the inflaton still dominates the Hubble expansion rate of the Universe, while transmitting its energy density to standard model (SM) radiation. Cosmic reheating ends when the SM energy density starts to dominate the total energy density of the Universe, at a temperature $\Trh$. If the (still unknown) value of $\Trh$ is much higher than the temperature at which DM is produced, the reheating dynamics decouples from the low-energy phenomenology. However, in the opposite case, the details of the reheating dynamics strongly impact the evolution of the DM abundance.

In this paper, we have considered the standard scenario in which a massive inflaton decays into SM particles with a constant decay width, injecting entropy into the SM bath. Additionally, we also took a simple extension of the SM with a $Z'$ gauge boson acting as a mediator and a fermionic DM. To follow the evolution of the background and the DM abundance, we implemented the required Boltzmann equations in the code micrOMEGAs to explore both the WIMP and the FIMP mechanisms. We determined the parameter space of the model that satisfies the relic density constraint under different assumptions for the inflaton decay width and examine current and future constraints from DM direct detection, taking special care of the scenarios where the DM was produced during the reheating era. We also determined the condition for DM to behave as a thermal or a nonthermal relic.

\acknowledgments
NB received funding from the Grant PID2023-151418NB-I00 funded by MCIU/AEI/10.13039 /501100011033/ FEDER, UE. GB and AP acknowledge the Department of Physics at NYU Abu Dhabi for their hospitality. The work of AP was carried out within the scientific program ``Particle Physics and Cosmology'' of the Russian National Center for Physics and Mathematics. This work was finalized when visiting the Institut Pascal at Université Paris-Saclay with the support of the program ``Investissements d'avenir'' ANR-11-IDEX-0003-01, the P2I axis of the Graduate School of Physics of Université Paris-Saclay, as well as IJCLab, CEA, IAS, OSUPS and APPEC.

\appendix
\section{Appendix} \label{sec:appendix}
New functions are available in micrOMEGAs to compute the DM relic density after solving the dynamics of the background given in Eqs.~\eqref{Infl_evol_rho} and~\eqref{Infl_evol_s}. We emphasize that we neglect DM production from inflaton decay. The function 

\verb|getInflDecay(HI, Gamma, &Trh, &Tmax, &aEnd)|\\
solves Eqs.~\eqref{Infl_evol_rho} and~\eqref{Infl_evol_s}.  The solution is stored in the tabulated functions Ta(a) and Ha(a), which respectively return the temperature and the Hubble parameter as a function of the scale parameter $a\in [1, \text{aEnd}]$. These functions are available to the user. Here, \verb|HI| is the Hubble rate at the end of inflation when HI is generated by the inflaton only. \verb|Gamma| is $\Gamma_\phi$, the decay width of the inflaton. \verb|Trh, Tmax| and \verb|aEnd| are return parameters. They represent respectively the temperature when the energy density of inflation becomes equal to the energy density of SM particles, the maximal temperature reached, and the expansion factor at the temperature \verb|Tend|. This temperature is defined by the user as a global parameter. The function \verb|getInflDecay| returns 0 when it can find a solution to the differential equations or 128 otherwise.
   
The function  

\verb|darkOmegaInfl(Beps, &isWimp, &err) |\\
solves the DM evolution equation in Eq.~\eqref{eq:BEc1}. It uses Ta and Ha constructed by \verb|getInflDecay|. The main return value of this routine is the DM relic density $\Omega_\chi h^2$. The auxiliary output parameters are {\it integer} {\tt isWIMP}, and {\it integer} {\tt err}. The parameter {\tt isWIMP} indicates whether the DM candidate behaves as a WIMP, that is, if the final relic density decreases when the cross section increases, then {\tt isWIMP=1}, otherwise {\tt isWIMP=0}. The error code {\tt err}  indicates whether there is a problem with the solution of the differential equations~\eqref{Infl_evol_s} and~\eqref{Infl_evol_rho}, in which case the 8$^\text{th}$ bit of the error code {\tt err} is 1, or with the solution of Eq.~\eqref{eq:BEc1}, then the 9$^\text{th}$ bit is 1. The lowest bits contain information about problems with numerical integration and should be treated as warnings. The parameter \verb|Beps| has the same meaning as in other \verb|darkOmega| routines; it determines the condition to include Boltzmann suppressed channels~\cite{Belanger:2001fz}.

The evolution of the DM abundance is stored in an array and is available via the functions {\tt Za(a)}, {\tt ZaEq(a)}, \verb|Ya(a)|, while the equilibrium abundance is stored in \verb|YaEq(a)|.

\bibliographystyle{JHEP}
\bibliography{biblio}

\providecommand{\href}[2]{#2}\begingroup\raggedright\begin{thebibliography}{10}

\bibitem{Bertone:2004pz}
G.~Bertone, D.~Hooper and J.~Silk, \emph{{Particle dark matter: Evidence,
  candidates and constraints}},
  \href{https://doi.org/10.1016/j.physrep.2004.08.031}{\emph{Phys. Rept.}
  {\bfseries 405} (2005) 279}
  [\href{https://arxiv.org/abs/hep-ph/0404175}{{\ttfamily hep-ph/0404175}}].

\bibitem{Cirelli:2024ssz}
M.~Cirelli, A.~Strumia and J.~Zupan, \emph{{Dark Matter}},
  \href{https://arxiv.org/abs/2406.01705}{{\ttfamily 2406.01705}}.

\bibitem{Planck:2018vyg}
{\scshape Planck} collaboration, \emph{{Planck 2018 results. VI. Cosmological
  parameters}},
  \href{https://doi.org/10.1051/0004-6361/201833910}{\emph{Astron. Astrophys.}
  {\bfseries 641} (2020) A6}
  [\href{https://arxiv.org/abs/1807.06209}{{\ttfamily 1807.06209}}].

\bibitem{Giudice:2000ex}
G.F.~Giudice, E.W.~Kolb and A.~Riotto, \emph{{Largest temperature of the
  radiation era and its cosmological implications}},
  \href{https://doi.org/10.1103/PhysRevD.64.023508}{\emph{Phys. Rev. D}
  {\bfseries 64} (2001) 023508}
  [\href{https://arxiv.org/abs/hep-ph/0005123}{{\ttfamily hep-ph/0005123}}].

\bibitem{Fornengo:2002db}
N.~Fornengo, A.~Riotto and S.~Scopel, \emph{{Supersymmetric dark matter and the
  reheating temperature of the universe}},
  \href{https://doi.org/10.1103/PhysRevD.67.023514}{\emph{Phys. Rev. D}
  {\bfseries 67} (2003) 023514}
  [\href{https://arxiv.org/abs/hep-ph/0208072}{{\ttfamily hep-ph/0208072}}].

\bibitem{Pallis:2004yy}
C.~Pallis, \emph{{Massive particle decay and cold dark matter abundance}},
  \href{https://doi.org/10.1016/j.astropartphys.2004.05.006}{\emph{Astropart.
  Phys.} {\bfseries 21} (2004) 689}
  [\href{https://arxiv.org/abs/hep-ph/0402033}{{\ttfamily hep-ph/0402033}}].

\bibitem{Gelmini:2006pw}
G.B.~Gelmini and P.~Gondolo, \emph{{Neutralino with the right cold dark matter
  abundance in (almost) any supersymmetric model}},
  \href{https://doi.org/10.1103/PhysRevD.74.023510}{\emph{Phys. Rev. D}
  {\bfseries 74} (2006) 023510}
  [\href{https://arxiv.org/abs/hep-ph/0602230}{{\ttfamily hep-ph/0602230}}].

\bibitem{Drees:2006vh}
M.~Drees, H.~Iminniyaz and M.~Kakizaki, \emph{{Abundance of cosmological relics
  in low-temperature scenarios}},
  \href{https://doi.org/10.1103/PhysRevD.73.123502}{\emph{Phys. Rev. D}
  {\bfseries 73} (2006) 123502}
  [\href{https://arxiv.org/abs/hep-ph/0603165}{{\ttfamily hep-ph/0603165}}].

\bibitem{Yaguna:2011ei}
C.E.~Yaguna, \emph{{An intermediate framework between WIMP, FIMP, and EWIP dark
  matter}}, \href{https://doi.org/10.1088/1475-7516/2012/02/006}{\emph{JCAP}
  {\bfseries 02} (2012) 006} [\href{https://arxiv.org/abs/1111.6831}{{\ttfamily
  1111.6831}}].

\bibitem{Roszkowski:2014lga}
L.~Roszkowski, S.~Trojanowski and K.~Turzy\'nski, \emph{{Neutralino and
  gravitino dark matter with low reheating temperature}},
  \href{https://doi.org/10.1007/JHEP11(2014)146}{\emph{JHEP} {\bfseries 11}
  (2014) 146} [\href{https://arxiv.org/abs/1406.0012}{{\ttfamily 1406.0012}}].

\bibitem{Drees:2017iod}
M.~Drees and F.~Hajkarim, \emph{{Dark Matter Production in an Early Matter
  Dominated Era}},
  \href{https://doi.org/10.1088/1475-7516/2018/02/057}{\emph{JCAP} {\bfseries
  02} (2018) 057} [\href{https://arxiv.org/abs/1711.05007}{{\ttfamily
  1711.05007}}].

\bibitem{Bernal:2018ins}
N.~Bernal, C.~Cosme and T.~Tenkanen, \emph{{Phenomenology of Self-Interacting
  Dark Matter in a Matter-Dominated Universe}},
  \href{https://doi.org/10.1140/epjc/s10052-019-6608-8}{\emph{Eur. Phys. J. C}
  {\bfseries 79} (2019) 99} [\href{https://arxiv.org/abs/1803.08064}{{\ttfamily
  1803.08064}}].

\bibitem{Bernal:2018kcw}
N.~Bernal, C.~Cosme, T.~Tenkanen and V.~Vaskonen, \emph{{Scalar singlet dark
  matter in non-standard cosmologies}},
  \href{https://doi.org/10.1140/epjc/s10052-019-6550-9}{\emph{Eur. Phys. J. C}
  {\bfseries 79} (2019) 30} [\href{https://arxiv.org/abs/1806.11122}{{\ttfamily
  1806.11122}}].

\bibitem{Cosme:2020mck}
C.~Cosme, M.~Dutra, T.~Ma, Y.~Wu and L.~Yang, \emph{{Neutrino portal to FIMP
  dark matter with an early matter era}},
  \href{https://doi.org/10.1007/JHEP03(2021)026}{\emph{JHEP} {\bfseries 03}
  (2021) 026} [\href{https://arxiv.org/abs/2003.01723}{{\ttfamily
  2003.01723}}].

\bibitem{Arias:2021rer}
P.~Arias, N.~Bernal, D.~Karamitros, C.~Maldonado, L.~Roszkowski and M.~Venegas,
  \emph{{New opportunities for axion dark matter searches in nonstandard
  cosmological models}},
  \href{https://doi.org/10.1088/1475-7516/2021/11/003}{\emph{JCAP} {\bfseries
  11} (2021) 003} [\href{https://arxiv.org/abs/2107.13588}{{\ttfamily
  2107.13588}}].

\bibitem{Bernal:2022wck}
N.~Bernal and Y.~Xu, \emph{{WIMPs during reheating}},
  \href{https://doi.org/10.1088/1475-7516/2022/12/017}{\emph{JCAP} {\bfseries
  12} (2022) 017} [\href{https://arxiv.org/abs/2209.07546}{{\ttfamily
  2209.07546}}].

\bibitem{Bhattiprolu:2022sdd}
P.N.~Bhattiprolu, G.~Elor, R.~McGehee and A.~Pierce, \emph{{Freezing-in
  hadrophilic dark matter at low reheating temperatures}},
  \href{https://doi.org/10.1007/JHEP01(2023)128}{\emph{JHEP} {\bfseries 01}
  (2023) 128} [\href{https://arxiv.org/abs/2210.15653}{{\ttfamily
  2210.15653}}].

\bibitem{Haque:2023yra}
M.R.~Haque, D.~Maity and R.~Mondal, \emph{{WIMPs, FIMPs, and Inflaton
  phenomenology via reheating, CMB and \ensuremath{\Delta}N$_{eff}$}},
  \href{https://doi.org/10.1007/JHEP09(2023)012}{\emph{JHEP} {\bfseries 09}
  (2023) 012} [\href{https://arxiv.org/abs/2301.01641}{{\ttfamily
  2301.01641}}].

\bibitem{Ghosh:2023tyz}
D.K.~Ghosh, A.~Ghoshal and S.~Jeesun, \emph{{Axion-like particle (ALP) portal
  freeze-in dark matter confronting ALP search experiments}},
  \href{https://doi.org/10.1007/JHEP01(2024)026}{\emph{JHEP} {\bfseries 01}
  (2024) 026} [\href{https://arxiv.org/abs/2305.09188}{{\ttfamily
  2305.09188}}].

\bibitem{Silva-Malpartida:2023yks}
J.~Silva-Malpartida, N.~Bernal, J.~Jones-P\'erez and R.A.~Lineros, \emph{{From
  WIMPs to FIMPs with low~reheating~temperatures}},
  \href{https://doi.org/10.1088/1475-7516/2023/09/015}{\emph{JCAP} {\bfseries
  09} (2023) 015} [\href{https://arxiv.org/abs/2306.14943}{{\ttfamily
  2306.14943}}].

\bibitem{Arias:2023wyg}
P.~Arias, N.~Bernal, J.K.~Osi\'nski, L.~Roszkowski and M.~Venegas,
  \emph{{Revisiting signatures of thermal axions in nonstandard cosmologies}},
  \href{https://doi.org/10.1103/PhysRevD.109.123529}{\emph{Phys. Rev. D}
  {\bfseries 109} (2024) 123529}
  [\href{https://arxiv.org/abs/2308.01352}{{\ttfamily 2308.01352}}].

\bibitem{Bernal:2024yhu}
N.~Bernal, K.~Deka and M.~Losada, \emph{{Thermal dark matter with
  low-temperature reheating}},
  \href{https://doi.org/10.1088/1475-7516/2024/09/024}{\emph{JCAP} {\bfseries
  09} (2024) 024} [\href{https://arxiv.org/abs/2406.17039}{{\ttfamily
  2406.17039}}].

\bibitem{Silva-Malpartida:2024emu}
J.~Silva-Malpartida, N.~Bernal, J.~Jones-P\'erez and R.A.~Lineros, \emph{{From
  WIMPs to FIMPs: Impact of Early Matter Domination}},
  \href{https://arxiv.org/abs/2408.08950}{{\ttfamily 2408.08950}}.

\bibitem{Bernal:2024ndy}
N.~Bernal, C.S.~Fong and {\'O}.~Zapata, \emph{{Probing low-reheating scenarios
  with minimal freeze-in dark matter}},
  \href{https://arxiv.org/abs/2412.04550}{{\ttfamily 2412.04550}}.

\bibitem{Sarkar:1995dd}
S.~Sarkar, \emph{{Big bang nucleosynthesis and physics beyond the standard
  model}}, \href{https://doi.org/10.1088/0034-4885/59/12/001}{\emph{Rept. Prog.
  Phys.} {\bfseries 59} (1996) 1493}
  [\href{https://arxiv.org/abs/hep-ph/9602260}{{\ttfamily hep-ph/9602260}}].

\bibitem{Kawasaki:2000en}
M.~Kawasaki, K.~Kohri and N.~Sugiyama, \emph{{MeV scale reheating temperature
  and thermalization of neutrino background}},
  \href{https://doi.org/10.1103/PhysRevD.62.023506}{\emph{Phys. Rev. D}
  {\bfseries 62} (2000) 023506}
  [\href{https://arxiv.org/abs/astro-ph/0002127}{{\ttfamily
  astro-ph/0002127}}].

\bibitem{Hannestad:2004px}
S.~Hannestad, \emph{{What is the lowest possible reheating temperature?}},
  \href{https://doi.org/10.1103/PhysRevD.70.043506}{\emph{Phys. Rev. D}
  {\bfseries 70} (2004) 043506}
  [\href{https://arxiv.org/abs/astro-ph/0403291}{{\ttfamily
  astro-ph/0403291}}].

\bibitem{DeBernardis:2008zz}
F.~De~Bernardis, L.~Pagano and A.~Melchiorri, \emph{{New constraints on the
  reheating temperature of the universe after WMAP-5}},
  \href{https://doi.org/10.1016/j.astropartphys.2008.09.005}{\emph{Astropart.
  Phys.} {\bfseries 30} (2008) 192}.

\bibitem{deSalas:2015glj}
P.F.~de~Salas, M.~Lattanzi, G.~Mangano, G.~Miele, S.~Pastor and O.~Pisanti,
  \emph{{Bounds on very low reheating scenarios after Planck}},
  \href{https://doi.org/10.1103/PhysRevD.92.123534}{\emph{Phys. Rev. D}
  {\bfseries 92} (2015) 123534}
  [\href{https://arxiv.org/abs/1511.00672}{{\ttfamily 1511.00672}}].

\bibitem{Allahverdi:2020bys}
R.~Allahverdi et~al., \emph{{The First Three Seconds: a Review of Possible
  Expansion Histories of the Early Universe}},
  \href{https://doi.org/10.21105/astro.2006.16182}{\emph{Open J.Astrophys.}
  {\bfseries 4} (2021) } [\href{https://arxiv.org/abs/2006.16182}{{\ttfamily
  2006.16182}}].

\bibitem{Batell:2024dsi}
B.~Batell et~al., \emph{{Conversations and Deliberations: Non-Standard
  Cosmological Epochs and Expansion Histories}},
  \href{https://arxiv.org/abs/2411.04780}{{\ttfamily 2411.04780}}.

\bibitem{Belanger:2018sti}
G.~B\'elanger et~al., \emph{{LHC-friendly minimal freeze-in models}},
  \href{https://doi.org/10.1007/JHEP02(2019)186}{\emph{JHEP} {\bfseries 02}
  (2019) 186} [\href{https://arxiv.org/abs/1811.05478}{{\ttfamily
  1811.05478}}].

\bibitem{Ahmed:2022tfm}
A.~Ahmed, B.~Grzadkowski and A.~Socha, \emph{{Higgs boson induced reheating and
  ultraviolet frozen-in dark matter}},
  \href{https://doi.org/10.1007/JHEP02(2023)196}{\emph{JHEP} {\bfseries 02}
  (2023) 196} [\href{https://arxiv.org/abs/2207.11218}{{\ttfamily
  2207.11218}}].

\bibitem{Cosme:2023xpa}
C.~Cosme, F.~Costa and O.~Lebedev, \emph{{Freeze-in at stronger coupling}},
  \href{https://doi.org/10.1103/PhysRevD.109.075038}{\emph{Phys. Rev. D}
  {\bfseries 109} (2024) 075038}
  [\href{https://arxiv.org/abs/2306.13061}{{\ttfamily 2306.13061}}].

\bibitem{Gan:2023jbs}
X.~Gan and Y.-D.~Tsai, \emph{{Cosmic Millicharge Background and Reheating
  Probes}},  \href{https://arxiv.org/abs/2308.07951}{{\ttfamily 2308.07951}}.

\bibitem{Cosme:2024ndc}
C.~Cosme, F.~Costa and O.~Lebedev, \emph{{Temperature evolution in the Early
  Universe and freeze-in at stronger coupling}},
  \href{https://doi.org/10.1088/1475-7516/2024/06/031}{\emph{JCAP} {\bfseries
  06} (2024) 031} [\href{https://arxiv.org/abs/2402.04743}{{\ttfamily
  2402.04743}}].

\bibitem{Arcadi:2024wwg}
G.~Arcadi, F.~Costa, A.~Goudelis and O.~Lebedev, \emph{{Higgs portal dark
  matter freeze-in at stronger coupling: observational benchmarks}},
  \href{https://doi.org/10.1007/JHEP07(2024)044}{\emph{JHEP} {\bfseries 07}
  (2024) 044} [\href{https://arxiv.org/abs/2405.03760}{{\ttfamily
  2405.03760}}].

\bibitem{Boddy:2024vgt}
K.K.~Boddy, K.~Freese, G.~Montefalcone and B.~Shams Es~Haghi, \emph{{Minimal
  Dark Matter Freeze-in with Low Reheating Temperatures and Implications for
  Direct Detection}},  \href{https://arxiv.org/abs/2405.06226}{{\ttfamily
  2405.06226}}.

\bibitem{Arcadi:2024obp}
G.~Arcadi, D.~Cabo-Almeida and O.~Lebedev, \emph{{$Z^\prime$-mediated dark
  matter freeze-in at stronger coupling}},
  \href{https://arxiv.org/abs/2409.02191}{{\ttfamily 2409.02191}}.

\bibitem{Lebedev:2024vor}
O.~Lebedev, A.P.~Morais, V.~Oliveira and R.~Pasechnik, \emph{{Invisible Higgs
  decay from dark matter freeze-in at stronger coupling}},
  \href{https://arxiv.org/abs/2410.21874}{{\ttfamily 2410.21874}}.

\bibitem{Alguero:2023zol}
G.~Alguero, G.~Belanger, F.~Boudjema, S.~Chakraborti, A.~Goudelis, S.~Kraml
  et~al., \emph{{micrOMEGAs 6.0: N-component dark matter}},
  \href{https://doi.org/10.1016/j.cpc.2024.109133}{\emph{Comput. Phys. Commun.}
  {\bfseries 299} (2024) 109133}
  [\href{https://arxiv.org/abs/2312.14894}{{\ttfamily 2312.14894}}].

\bibitem{Drees:2015exa}
M.~Drees, F.~Hajkarim and E.R.~Schmitz, \emph{{The Effects of QCD Equation of
  State on the Relic Density of WIMP Dark Matter}},
  \href{https://doi.org/10.1088/1475-7516/2015/06/025}{\emph{JCAP} {\bfseries
  06} (2015) 025} [\href{https://arxiv.org/abs/1503.03513}{{\ttfamily
  1503.03513}}].

\bibitem{BICEP:2021xfz}
{\scshape BICEP, Keck} collaboration, \emph{{Improved Constraints on Primordial
  Gravitational Waves using Planck, WMAP, and BICEP/Keck Observations through
  the 2018 Observing Season}},
  \href{https://doi.org/10.1103/PhysRevLett.127.151301}{\emph{Phys. Rev. Lett.}
  {\bfseries 127} (2021) 151301}
  [\href{https://arxiv.org/abs/2110.00483}{{\ttfamily 2110.00483}}].

\bibitem{Barman:2021ugy}
B.~Barman and N.~Bernal, \emph{{Gravitational SIMPs}},
  \href{https://doi.org/10.1088/1475-7516/2021/06/011}{\emph{JCAP} {\bfseries
  06} (2021) 011} [\href{https://arxiv.org/abs/2104.10699}{{\ttfamily
  2104.10699}}].

\bibitem{Cosme:2021baj}
C.~Cosme, M.~Dutra, S.~Godfrey and T.R.~Gray, \emph{{Testing freeze-in with
  axial and vector Z' bosons}},
  \href{https://doi.org/10.1007/JHEP09(2021)056}{\emph{JHEP} {\bfseries 09}
  (2021) 056} [\href{https://arxiv.org/abs/2104.13937}{{\ttfamily
  2104.13937}}].

\bibitem{Griest:1989wd}
K.~Griest and M.~Kamionkowski, \emph{{Unitarity Limits on the Mass and Radius
  of Dark Matter Particles}},
  \href{https://doi.org/10.1103/PhysRevLett.64.615}{\emph{Phys. Rev. Lett.}
  {\bfseries 64} (1990) 615}.

\bibitem{Bhatia:2020itt}
D.~Bhatia and S.~Mukhopadhyay, \emph{{Unitarity limits on thermal dark matter
  in (non-)standard cosmologies}},
  \href{https://doi.org/10.1007/JHEP03(2021)133}{\emph{JHEP} {\bfseries 03}
  (2021) 133} [\href{https://arxiv.org/abs/2010.09762}{{\ttfamily
  2010.09762}}].

\bibitem{Bernal:2023ura}
N.~Bernal, P.~Konar and S.~Show, \emph{{Unitarity bound on dark matter in
  low-temperature reheating scenarios}},
  \href{https://doi.org/10.1103/PhysRevD.109.035018}{\emph{Phys. Rev. D}
  {\bfseries 109} (2024) 035018}
  [\href{https://arxiv.org/abs/2311.01587}{{\ttfamily 2311.01587}}].

\bibitem{LZCollaboration:2024lux}
{\scshape LZ Collaboration} collaboration, \emph{{Dark Matter Search Results
  from 4.2 Tonne-Years of Exposure of the LUX-ZEPLIN (LZ) Experiment}},
  \href{https://arxiv.org/abs/2410.17036}{{\ttfamily 2410.17036}}.

\bibitem{DARWIN:2016hyl}
{\scshape DARWIN} collaboration, \emph{{DARWIN: towards the ultimate dark
  matter detector}},
  \href{https://doi.org/10.1088/1475-7516/2016/11/017}{\emph{JCAP} {\bfseries
  11} (2016) 017} [\href{https://arxiv.org/abs/1606.07001}{{\ttfamily
  1606.07001}}].

\bibitem{Belanger:2001fz}
G.~B\'elanger, F.~Boudjema, A.~Pukhov and A.~Semenov, \emph{{MicrOMEGAs: A
  Program for calculating the relic density in the MSSM}},
  \href{https://doi.org/10.1016/S0010-4655(02)00596-9}{\emph{Comput. Phys.
  Commun.} {\bfseries 149} (2002) 103}
  [\href{https://arxiv.org/abs/hep-ph/0112278}{{\ttfamily hep-ph/0112278}}].

\end{thebibliography}\endgroup
\end{document}